\documentclass[preprint,prl,aps,amsmath,amssymb,color,superscriptaddress,floatfix]{revtex4}
\usepackage{stmaryrd}
\usepackage{amsmath}
\usepackage{amssymb}
\usepackage{graphicx}
\usepackage{dcolumn}
\usepackage{bm}
\usepackage{tabularx}
\usepackage{color}
\usepackage[normalem]{ulem}
\begin{document}
\begin{titlepage}
		\title{Are there type-III multiferroics?}
		
		\author{Haojin Wang}
		\affiliation{Key Lab of advanced optoelectronic quantum architecture and measurement (MOE), and Advanced Research Institute of Multidisciplinary Science, Beijing Institute of Technology, Beijing 100081, China}
		\author{Haitao Liu}
		\affiliation {Institute of Applied Physics and Computational Mathematics, Beijing 100088, China}
		\affiliation {National Key Laboratory of Computational Physics, Beijing 100088, China}
		\author{Meng Ye}
		\email{mye@gscaep.ac.cn}
		\affiliation {Graduate School of China Academy of Engineering Physics, Beijing 100193, China}
		\author{Yuanchang Li}
		\email{yuancli@bit.edu.cn}
		\affiliation{Key Lab of advanced optoelectronic quantum architecture and measurement (MOE), and Advanced Research Institute of Multidisciplinary Science, Beijing Institute of Technology, Beijing 100081, China}
		
		\date{\today}
		
\begin{abstract}
Multiferroics are known to be classified into two types. However, type-I lacks sufficient magnetoelectric coupling and type-II lacks sufficient electric polarization, making both practically difficult. In this work, we explore the possibility of type-III multiferroics, where the origins of ferroelectricity and magnetism are highly intertwined but not causally related, with a combination of strong magnetoelectric coupling and large polarization. Our first-principles calculations predict that monolayer TiCdO$_{4}$ is such a type-III ferroelectric-ferromagnetic multiferroics with both electronic and magnetic orders originating from competing electron populations on oxygen atoms. It shows an electric polarization of 50 $\mu$C/m$^{2}$ while the maximum linear and quadratic magnetoelectric response are as high as 35000 ps/m and 1.59 $\times$ 10$^{-14}$ s/A, respectively. Our study opens up new perspectives for the discovery and design of much-anticipated multiferroics that can be used for cross-modulation.
\end{abstract}
		
		\maketitle
		\draft
		\vspace{2mm}
\end{titlepage}

Magnetoelectric multiferroics, which integrates ferroelectricity and magnetism, present remarkable potential for application in memory, data processing, and energy harvesting/conversion devices\cite{mu4,muti}. These materials offer a cross-coupling pathway that enables electrically controlled magnetism or magnetically controlled ferroelectricity. Single-phase multiferroics are classified into two types based on the correlation between the ferroelectric and magnetic origins\cite{Khomskii}. For type-I, the origins are mutually independent, leading to weak magnetoelectric coupling. A paradigmatic instance is BiFeO$_3$, where ferroelectricity arises from the lone-pair electrons on Bi (6$s^2$) and magnetism originates from Fe-O hybridization\cite{bfo,Ravindran}. Despite having a large polarization of 50$\sim$60 $\mu$C/cm$^2$, comparable to conventional ferroelectrics such as BaTiO$_3$\cite{Merz} and PbTiO$_3$\cite{Nishino}, its magnetoelectric coupling is limited to approximately 1 ps/m\cite{Ederer,muti}.

In contrast, type-II multiferroics derive their ferroelectricity from spin-induced inversion symmetry breaking, leading to strong magnetoelectric coupling, exemplified by TbMnO$_3$ ($\sim$250 ps/m)\cite{tmo} and TbMn$_2$O$_5$ ($\sim$500 ps/m)\cite{Hur}. However, due to the dependence of ferroelectricity on magnetic ordering, the polarization is usually small, around 0.01 $\mu$C/cm$^2$, which is only one-thousandth of that observed in type-I multiferroics\cite{Zhai,Xiang,type2-3,type2-4,type2-5}. For practical applications, an ideal multiferroic material should have both a large polarization and a strong magnetoelectric coupling, yet neither type-I nor type-II can meet this requirement\cite{Dong}. Consequently, there is a demand for a new type of multiferroics wherein the ferroelectric and magnetic origins are intricately intertwined, yet not causally related.
	
The pursuit of new multiferroics is also of great scientific interest. In quantum mechanics, charge and spin are two sides of a coin for electrons, and their mutual interactions rely on spin-orbit coupling. Ferroelectricity and magnetism are macroscopic charge and spin orders that break space-inversion and time-reversal symmetries, respectively. What kind of microscopic mechanism enables a strong coupling between the two macroscopic ferro-orders? From an academic point of view, the discovery of type-II multiferroics is a milestone in modern physics that has greatly advanced the fundamental understanding of the interplay between charge and spin orders.
	
In this work, first-principles calculations have identified monolayer TiCdO$_{4}$ as a multiferroic that does not fall into either type-I or type-II classifications. It features simultaneous $d^0$-ferroelectricity and $d^0$-ferromagnetism, both of which arise from competing electron populations on the four O atoms. As a result, on the one hand, it shows an electric polarization of 50 $\mu$C/cm$^2$, which is comparable to that of type-I multiferroics. On the other hand, the electron-population fluctuations on the O-atoms mediate a giant nonlinear magnetoelectric coupling, with maximum linear and quadratic responses reaching record $\alpha_{yx}$ = 35000 ps/m and $\beta_{yzz}$ = 1.59 $\times 10^{-14}$ s/A. Based on these findings, we propose the type-III multiferroics, whose ferroelectric and magnetic origins are highly entangled, along with large electric polarization and strong magnetoelectric coupling. Our findings provide new insights into the search for and design of multiferroics with practical value, and thus ultimately magnetoelectric cross-modulation.

\begin{figure*}[htp]
	\centering
	\begin{center}
		\includegraphics[width=1\columnwidth]{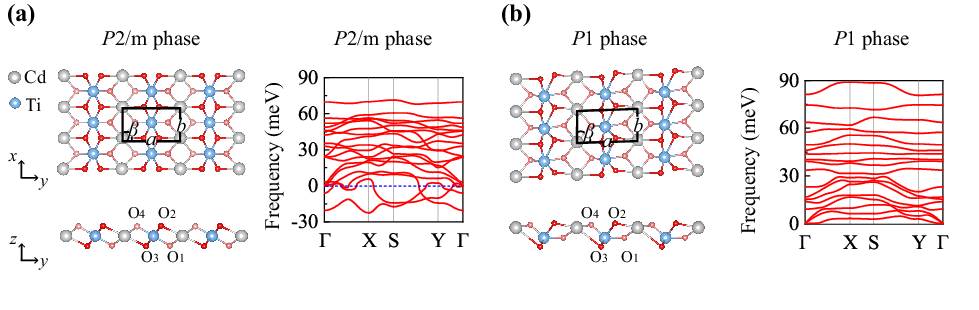}
		\caption{\label{fig:fig1} Geometric configuration and the corresponding phonon spectrum of (a) symmetric $\textit{P}2/m$ and (b) ferroelectric $\textit{P}1$ TiCdO$_{4}$ monolayer. A unit cell (black tetragonal) contains 4 O-atoms. The symmetric phase is centrosymmetric with respect to the metal atoms. Consequently, O$_1$ and O$_4$ are equivalent and are bonded to two Cd atoms and one Ti atom, while O$_2$ and O$_3$ are equivalent and are bonded to two Ti atoms and one Cd atom. Although the angle $\beta$ deviates slightly from 90$^{\circ}$ in the ferroelectric phase, for the sake of comparison, we have followed the same Brillouin-zone path as the symmetric phase throughout this work.}
		\label{fig:f1}
	\end{center}
\end{figure*}	
		
Density functional theory calculations were performed using the Vienna Ab initio Simulation Package (VASP)\cite{vasp} with Perdew-Burke-Ernzerhof (PBE)\cite{pbe} exchange-correlation functional and projector-augmented-wave (PAW) approach\cite{paw,paww}. To ensure the findings, the HSE06 functional\cite{hse} was further employed for structural relaxations and electronic structure calculations. The plane-wave cutoff energy was set to 600 eV. A 24 $\times$ 15 $\times$ 1 Monkhorst-Pack $k$-mesh was used to sample the Brillouin zone. An 18 \AA\ vacuum layer was added to minimize spurious interactions between the monolayer and its periodic images. Van der Waals interactions were included in Grimme's DFT-D3 scheme\cite{vdw}. Lattice vectors and atomic positions were fully relaxed until the total energy and residual force converged to within 10$^{-6}$ eV and 10$^{-3}$ eV/\AA, respectively. The phonon spectrum was computed using the finite displacement method in the PHONOPY code\cite{phonon}, which employs 4 $\times$ 4 $\times$ 1 supercells. Polarization was calculated within the Berry phase formalism\cite{berry}. The magnetoelectric response was obtained from first-principles calculations including spin-orbit coupling by tracking the polarization change under a uniform Zeeman field\cite{zeeman1}. This scheme has been successfully applied to study the magnetoelectric effects of hexagonal rare-earth manganites and ferrites\cite{Ye2014,Ye2015}.	

Figure 1(a) presents the geometrical configuration of the high-symmetry TiCdO$_{4}$ monolayer, which can be viewed as half of the Ti in 1$T$-TiO$_{2}$ being substituted by Cd\cite{bu-prb}. Although this substitution decreases the space group from $P\overline{3}$m1 to $P$2/m, TiCdO$_{4}$ still retains an inversion symmetry centered on the transition-metal atoms. However, the corresponding phonon spectrum exhibits significant imaginary frequencies, suggesting that the structure is thermodynamically unstable.

A stable low-symmetry TiCdO$_{4}$ is obtained after fully relaxing the lattice vectors and atomic positions. Its crystal structure and phonon spectrum are shown in Fig. 1(b). Energetically, the spontaneous distortion reduces the system energy by 276 meV per formula. Structurally, the lattice constant $a$($b$) shrinks (stretches) by 0.1(0.2) \AA\ and the angle $\beta$ changes from 90 to 88$^{\circ}$. The system no longer has any spatial symmetry and the space group is $P$1. Deformation in TiCdO$_{4}$ differs very much from the case of early-transition-metal substitution\cite{wang-jmcc}, where there is a significant out-of-plane displacement of the metal and a negligible displacement of skeleton O atoms. In contrast, both Ti/Cd and O atoms undergo large displacements in TiCdO$_{4}$.

Inversion symmetry breaking gives an out-of-plane polarization of 3 $\mu$C/cm$^{2}$ and an in-plane polarization of 50 $\mu$C/cm$^{2}$ along the ($-$0.94, $-$0.34, 0) direction\cite{note}. The polarization strength is comparable to that of type-I multiferroics such as BiFeO$_{3}$\cite{bfo}, YMnO$_{3}$\cite{ymno3} and YbFeO$_{3}$\cite{ybfeo3}.

The stability of TiCdO$_{4}$ is further evaluated by estimating its formation energy using the formula
\begin{equation}\label{(1)}
		E_{\rm f}= (E_{\rm {TiCdO_4}}-\mu_{\rm {Ti}}-\mu_{\rm {Cd}}-2\mu_{\rm {O_2}})/6,
\end{equation}
where $E_{\rm {TiCdO_4}}$ is the total energy of distorted TiCdO$_{4}$. $\mu _{\rm Ti}$/$\mu _{\rm Cd}$ and $\mu_{\rm O}$ are the chemical potentials of hexagonal Ti/Cd crystals and gas-phase O$_{2}$, respectively. It gives $E_{\rm f}$ = $-$1.44 eV. A negative $E_{\rm f}$ indicates that the production of TiCdO$_{4}$ is exothermic and the reaction can proceed spontaneously.

\begin{figure*}[tbp]
		\centering
		\includegraphics[width=1\columnwidth]{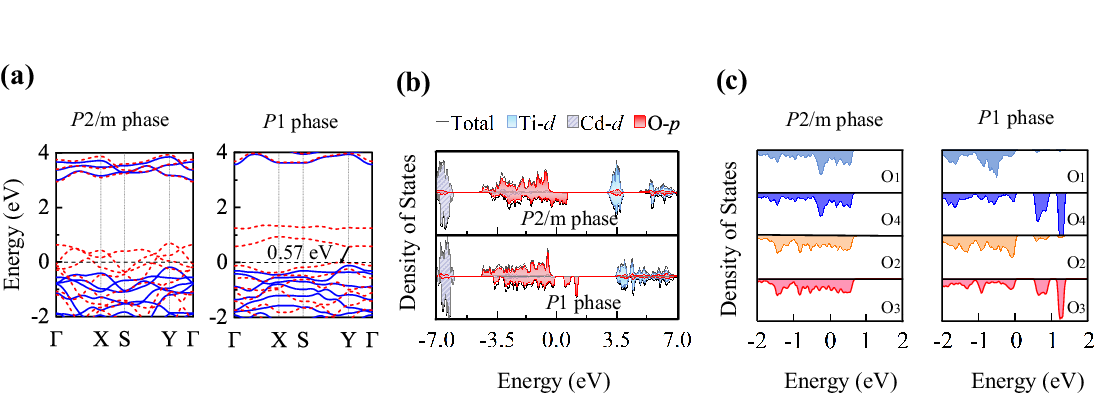}
		\caption{\label{fig:fig2}  (a) Spin-resolved band structures of symmetric ($\textit{P}2/m$) and ferroelectric ($\textit{P}1$) TiCdO$_{4}$ monolayer. Blue solid and red dashed lines indicate the spin-majority and spin-minority channels, respectively. The black arrow denotes the minimum gap. (b) Total and projected density of states (PDOS) of symmetric (Top) and ferroelectric (Bottom) TiCdO$_{4}$. (c) PDOS to the four O-atoms (See Fig. 1) in the spin-minority channel for symmetric and ferroelectric phase, respectively. The Fermi levels are set to energy zero.
		}
\end{figure*}

Figure 2(a) compares the spin-resolved band structures of symmetric and ferroelectric TiCdO$_{4}$. The former is a half-metal\cite{YCL} with a spin-majority gap of 3.19 eV and a spin-minority metal. The latter is a half-semiconductor\cite{halfSemi} with a spin-majority gap of 3.67 eV and a spin-minority gap of 0.57 eV. Evidently, spontaneous polarization is accompanied by a spin-minority metal-insulator transition.

Electronic structure calculations give a total magnetic moment of 2 $\mu_{\rm B}$ for both symmetric and ferroelectric TiCdO$_{4}$. The projected density of states depicted in Fig. 2(b) show that O-2$p$ always dominates near the Fermi energy, while the Ti-3$d$ and Cd-4$d$ states are located deeper in the conduction and valence bands, respectively. It is clear that the magnetic moments come mainly from the O-2$p$ states. As a result, TiCdO$_{4}$ displays unconventional $d^0$-magnetism\cite{d0mag-2005,BN-prl,zno-prl}, which has been extensively studied by the magnetism community, but has long been overlooked by the multiferroics community.

If one treats each unit cell as a magnetic entity, an energy comparison of the four different magnetic configurations depicted in Fig. S1\cite{SI} yields that both symmetric and ferroelectric TiCdO$_{4}$ have a ferromagnetic ground state. The ferroelectric TiCdO$_{4}$ has an in-plane easy magnetization axis with a magnetic anisotropy energy of 68 $\mu$eV per formula with respect to the out-of-plane hard axis. This value is an order of magnitude higher than 1.4 $\mu$eV of Fe\cite{fe-prb}, suggesting that TiCdO$_{4}$ can overcome the thermal fluctuations and maintain long-range order.

We have carried out further structural relaxation and electronic structure calculations using the more accurate HSE06 functional to make sure that the discovery of ferroelectric-ferromagnetic multiferroics is not due to PBE. The results show that this holds true. The physics including spontaneous distortion, polarization and spin-minority metal-insulator transition remain unchanged (See the Supplementary Material for details\cite{SI}). The O-2$p$ states still dominate around the Fermi energy and contribute to the $d^0$-magnetism. The changes by HSE06 are quantitative. For example, the spin-minority gap of distorted TiCdO$_{4}$ increases from 0.57 eV of PBE to 2.19 eV. Therefore, we continue our discussion based on PBE in the following.

\begin{figure}[tp]
		\begin{center}
			\includegraphics[width=0.99\columnwidth]{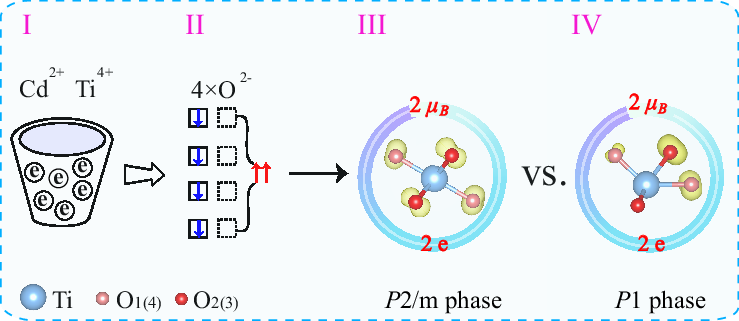}
			\caption{\label{fig:fig3} Schematic diagram of TiCdO$_4$ multiferroics induced by competing-electron-populations among O-atoms. In a unit cell, the metal cations (Ti and Cd) provide a total of 6 electrons (I), which is not sufficient for all 4 O-atoms to become stable $O^2$. The four spin-minority orbitals have to ``compete" for 2 electrons (II), leading to a total magnetic moment of 2 $\mu_{\rm B}$ that is almost independent of the atomic configurations. However, the number of electrons ``captured" by each O-atom determines its local moment, which depends on the atomic configurations, e.g., symmetric (III) and ferroelectric (IV) cases. On the other hand, the partial occupation of four orbitals of similar energy by 2 electrons triggers a Jahn-Teller-like instability, which changes the metal-O bond length and bond angle. Induced ionic shifts break the centrosymmetry with respect to the metal atoms, resulting in spontaneous ferroelectric polarization. Therefore, both the ferroelectricity and magnetism of TiCdO$_4$ originate from competing-electron-populations among O-atoms. Application of an external magnetic field would cause a population fluctuation, which alters the metal-O bonding and directly affects the electric polarization. Likewise, the application of an external electric field would change the electron population among O-atoms, which alters the local moment on each O-atom (Yellow dots in III and IV denote spin density isosurface of 0.02 e/\AA$^{3}$) and the direction of the easy magnetization axis. As a result, TiCdO$_4$ exhibits strong magnetoelectric coupling that is largely independent of spin-orbit interaction.}
			\end{center}
\end{figure}

Unambiguously, TiCdO$_{4}$ is not a type-II multiferroic because its ferroelectricity arises from ionic displacements rather than spin structure. Unlike the usual displacive ferroelectrics where only the metal cations move off-center\cite{Tan}, here the anion O also shifts significantly reminding its key role in the structural distortion. Given that the spin-minority metal-insulator transition is also O-dominated, we further compare the spin-minority density of states of the four O atoms before and after the distortion in Fig. 2(c). A notable change is that the distortion leads to an accumulation of electrons towards O$_1$ and O$_2$, making the two originally equivalent O atoms [O$_1$ = O$_4$ and O$_2$ = O$_3$, see also Fig. 1(a)] no longer identical.

Figure 3 illustrates what we call a competing-electron-population mechanism to explain the multiferroics of TiCdO$_{4}$. Ti/Cd has a valence electron configuration of 3$d^2$4$s^2$/4$d^{10}$5$s^2$. When crystallized into TiCdO$_{4}$, they lose 4 and 2 valence electrons to form Ti$^{4+}$ and Cd$^{2+}$, respectively. For the four O-atoms, a total of eight electrons are required to make them all stable O$^{2-}$. In the ``electron-deficient" environment where the metal provides only 6 electrons, the spin-majority orbitals of O are completely filled but the four spin-minority orbitals have to compete for the remaining 2 electrons. This results in a total magnetic moment of 2 $\mu_B$ that is almost independent of atomic configuration. Yet, the local moment of each O-atom depends on the number of electrons it ``captures". The ``competitive population" of 2 electrons on four spin-minority orbitals of similar energy in turn triggers a Jahn-Teller-like instability. This leads to a change in the metal-O bonding (bond lengths and angles), which drives the ions to move and breaks the inversion symmetry. Therefore, both the ferroelectric and magnetic orders of TiCdO$_{4}$ originate from the competing-electron-population on the O-atoms, making it not a type-I multiferroic either.

As the ferroelectricity of TiCdO$_{4}$ is $d$-electron independent, it is not surprising that it has the large polarization of conventional $d^0$-ferroelectrics. Meanwhile, high entanglement of the ferroelectric and magnetic orders naturally hints a strong magnetoelectric coupling. This is because applying an electric/magnetic field would induce fluctuations in the electron populations among the O-atoms, consequently changing the magnetic/electrical properties. To account for this quantitatively, we directly calculate the polarizations $\textbf{\textit{P}}$ under different applied Zeeman fields $\textbf{\textit{H}}$ and the results are summarized in Fig. S3\cite{SI}. Then, we use
	\begin{eqnarray}
		P_{i} = \alpha_{ij}H_{j}+\frac{1}{2}\beta_{ijj}H_{j}H_{j}+ \cdot \cdot \cdot
	\end{eqnarray}
to fit the linear and quadratic magnetoelectric coupling tensors
\begin{equation}
\bm{\alpha} =
\begin{pmatrix}
  8540  & -3470 &  -610  \\
-35000  &  2960 &  6330 \\
 -1620  &  9130 &  2630 \\
\end{pmatrix}
ps/m,
\end{equation}
and
\begin{equation}
\bm{\beta} =
\begin{pmatrix}
 -0.24   &  0.45 &  0.33  \\
  1.34   & -1.37 & -1.59  \\
  0.06   & -0.61 & -0.13  \\
\end{pmatrix}
\times 10^{-14} s/A.
\end{equation}

\begin{figure}[tbp]
		\begin{center}
			\includegraphics[width=0.8\columnwidth]{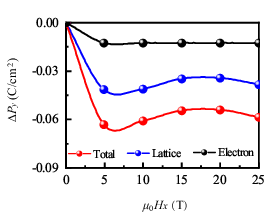}
			\caption{\label{fig:fig4} Total variation of the polarization in the $y$ direction ($\Delta P_y$) with applied magnetic fields in the $x$ direction ($H_x$), and decomposition into separate contributions from the lattice and electrons. The points are calculated directly from first-principles (See the Supplementary Material for details), while the lines are guides to the eye.
			}
		\end{center}
\end{figure}

One can see that the minimum linear response $\mid\alpha_{xz}\mid$ = 610 ps/m is larger than that of type-II multiferroics TbMnO$_3$\cite{tmo} and TbMn$_2$O$_5$\cite{Hur}. The maximum of $\mid\alpha_{yx}\mid$ = 35000 ps/m is a record, larger than other strong magnetoelectric multiferroics known to date, including Ba$_{0. 4}$Sr$_{1.6}$Mg$_{2}$Fe$_{12}$O$_{22}$ (33000 ps/m)\cite{Zhai}, Fe$_{2}$Mo$_{3}$O$_{8 }$ (10000 ps/m)\cite{fe2mo3o8-sci.rep} and Ni$_{3}$TeO$_{6}$ (1300 ps/m)\cite{ni3teo6-nc}. The maximum quadratic response $\mid\beta_{yzz}\mid$ = 1.59 $\times$10$^{-14}$ s/A is also two orders of magnitude larger than that of CaMnO$_{3}$\cite{camno3-prl}. It is worth noting that the $\textbf{\textit{P}}$-$\textbf{\textit{H}}$ relation calculated without spin-orbit coupling, as well as the derived $\bm{\alpha}$ and $\bm{\beta}$, remain almost unchanged. This implies that the microscopic mechanism dominating the giant magnetoelectric coupling in TiCdO$_{4}$ is not spin-orbit coupling, which indirectly supports the electron-population-fluctuation mechanism shown in Fig. 3.

Typically, the linear magnetoelectric response includes contributions from electrons ($\alpha_{ij}^{e}$) and lattice ($\alpha_{ij}^{l}$)\cite{Iniguez}. In order for a deeper understanding of the magnetoelectric coupling in TiCdO$_{4}$, we calculate $\alpha_{ij}^{e}$ and $\alpha_{ij}^{l}$ separately (See the Supplementary Material for details\cite{SI}). With $\mid\alpha_{yx}\mid$ maximal, we give representatively in Fig. 4 the total and decomposed changes in $P_y$ under the application of $H_x$. It can be seen that the $\alpha_{yx}^{l}$ is approximately three times that of the $\alpha_{yx}^{e}$. Moreover, the $\alpha_{ij}^{l}$ can be decomposed as\cite{Ye2015}
\begin{eqnarray}
	\alpha_{ij}^{l} = \frac{\partial P_{i}}{\partial u_{m}}\cdot\frac{\partial u_{m}}{\partial F_{n}}\cdot\frac{\partial F_{n}}{\partial H_{j}}=\Omega_{0}^{-1}Z^{e}_{mi}\cdot K^{-1}_{mn} \cdot \mu_{0}Z^{m}_{nj},
\end{eqnarray}
where $m$ and $n$ are directions, and $\Omega_{0}$ is the unit cell volume. The three terms on the right-hand side measure the ability to induce polarization with the same ionic displacement, the ability to induce ionic displacement with the same force, and the ability to generate force with the same magnetic field. Traditionally, they are notated as the Born effective charge ($Z^e$), the inverse of the force constant ($K^{-1}$) and the magnetic charge ($Z^m$). Compared to the prototypical multiferroic Cr$_{2}$O$_{3}$\cite{cr2o3,zeeman1}, the $\mid\alpha_{yx}\mid$ of TiCdO$_{4}$ increases by about four orders of magnitude. Our decomposition calculations of Eq. (5) show that $Z^e$ is almost unchanged, $K^{-1}$ increases by a factor of four, and $Z^m$ increases by three orders of magnitude. Unambiguously, the giant magnetoelectric response of TiCdO$_{4}$ originates from its huge $Z^m$.

In summary, we have identified by first-principles calculations that the monolayer TiCdO$_{4}$ is a ferroelectric-ferromagnetic multiferroics, which does not belong to the traditional type-I or type-II classification. Its ferroelectric and magnetic origins are highly intertwined, both originating from the competing-electron-population among O-atoms under an ``electron-deficient" environment. This enables an integration of the large electric polarization of type-I and the strong magnetoelectric coupling of type-II, two necessary requirements for the multiferroic applications. We therefore classify it as a type-III multiferroic, which is expected to facilitate the search for and design of multiferroics that can ultimately achieve cross-control between ferroelectricity and magnetism.

Some useful insights can be gained from TiCdO$_{4}$ about which materials might be type-III multiferroics. It combines $d^0$-ferroelectricity and $d^0$-magnetism. It is well known that $d^0$-ferroelectrics are sufficiently polarized, and that $d^0$-magnetism emerges when the 2$p$ orbitals of C, N or O are not completely filled. Since the 2$p$ orbitals are more extended than the 3$d$ ones, they behave more ``softly" in the presence of an external field (e.g., TiCdO$_{4}$ under a magnetic field and organometallic wires under an electric field\cite{LiuJ}), resulting in a larger response. In addition, there is usually more than one anion in the unit cell, so $d^0$-magnet can host an internal magnetic structure. Modulation of this magnetic structure by an external field can produce a rich variety of magnetoelectric effects.

This work was supported by the Ministry of Science and Technology of China (Grant Nos. 2023YFA1406400 and 2020YFA0308800), the National Natural Science Foundation of China (Grant No. 12474064), and NSAF (Grant No. U2330401).


\begin{thebibliography}{90}
\bibitem{note} Given that the high-symmetry phase exhibits metallicity, the Berry phase calculations were performed using the QUANTUM ESPRESSO\cite{qe} package. Meanwhile, the thickness of the monolayer is assumed to be 2 \AA.
\end{thebibliography}
\end{document}